\documentclass[
onecolumn,%
nobibnotes,%
nofootinbib,
eqsecnum,showpacs,amsmath,amssymb,onecolumn]{revtex4}
\usepackage{graphicx}%
\usepackage{amsmath}
\usepackage{amsfonts}%
\usepackage{amssymb}
\begin{document}
\title{Pion radii in nonlocal chiral quark model}
\author{A.~E.~Dorokhov}
\author{A.~E.~Radzhabov}
\author{M.~K.~Volkov}

\affiliation{Bogoliubov Laboratory of Theoretical Physics, \\
Joint Institute for Nuclear Research, 
141980, Dubna, Russia}

\begin{abstract}
The electromagnetic radius of the charged pion and the transition radius of the
neutral pion are calculated in the framework of the nonlocal chiral quark model.
It is shown in this model that the contributions of vector mesons to the pion radii
are noticeably suppressed in comparison
with a similar contribution in the local Nambu--Jona-Lasinio model.
The form-factor for the process $\gamma^*\pi^+ \pi^-$ is calculated for the $-1$ GeV$^2$$<q^2<1.6$ GeV$^2$.
Our results are in satisfactory agreement with experimental data.
\end{abstract}
\pacs{14.40.-n,11.10.Lm,12.39.Ki}
\maketitle

\section{Introduction}
In our works \cite{DRV,RV} the nonlocal chiral quark model with quark
form-factors of the gaussian type was proposed. This approach is the
development of the nonlocal quark models considered in
\cite{Bowler:ir,DoLT98,Dorokhov:2001wx}. These models are nonlocal
extensions of the well-known Nambu--Jona-Lasinio (NJL) 
model with local quark interaction \cite{Volkov:zb,ZC}. 
However, with nonlocal form-factors, quark loops are free of
ultraviolet (UV) divergences and the quark confinement is possible to implement.
Moreover, the nonlocal structure may be motivated \cite{DoLT98} by fundamental
QCD interactions induced by the instanton and gluon exchanges
\cite{Sh82,Diakonov:1983hh,DK92}, which leads to spontaneous breaking of the
chiral symmetry, solves $U_{A}\left(  1\right)  $ problem and dynamically
generates a momentum-dependent quark mass. The use of a covariant nonlocal
low-energy quark model based on a self-consistent approach to the dynamics
of quarks has many attractive features: it preserves gauge invariance,
it is consistent with the low-energy theorems as well as takes into account the
large-distance dynamics controlled by bound states
\cite{Bowler:ir,DoLT98,DB03,Do03pisma}. Masses and strong decays of the
scalar, vector and axial-vector mesons were considered earlier in
\cite{DRV,RV,Bowler:ir}.

Nonlocal models, in contrast with the local NJL model, can be successfully
used for the description of not only the constant part of amplitudes of meson
interactions but also of the momentum expansion of amplitudes at small
energies. With the help of these expansions it is possible to describe a set of
important meson properties: electromagnetic radii, electric and magnetic
polarizabilities, meson-meson scattering lengths, slope parameters and so on.

In the standard local NJL models \cite{Volkov:zb,ZC} such expansions are not
valid and can lead to incorrect results. In order to demonstrate this, let us
consider the calculation of the pion radius. The expansion of the triangle
quark diagram (fig. \ref{pchrad}a) with two pions and one photon connected
locally to quark lines\footnote{In the following this kind of diagrams will be
called contact diagrams.} leads to satisfactory agreement with experimental
data \cite{Gerasimov,Hippe:hu,Volkov:1996rc}. However in addition, it is
necessary to take into account contribution of the diagram with the
intermediate $\rho$-meson, fig. \ref{pchrad}e. In the framework of the local
NJL model this contribution is comparable with that of the contact diagram. As
a result, we obtain a too large value for the pion radius which is in
disagreement with experimental data \cite{Volkov:1996rc}. Some authors
attempted to solve this problem, using different approaches. Some of them
ignored this contribution \cite{Hippe:hu,Tarrach:ta} while the others used
special methods beyond the standard local NJL model
\cite{Volkov:1996rc,Bernard:1988bx,Lutz:zc}.

In this work, we demonstrate that this problem can be solved in the framework
of the nonlocal models without any additional assumption. In these models,
the contribution of the diagrams with intermediate vector mesons is shown to
be suppressed. It is essentially smaller than the contribution of the contact
diagram. However, without $\rho$-meson diagrams it is impossible to describe
form-factor of the process $\gamma^{\ast}\pi^{+}\pi^{-}$ in the time-like region.
Our results are in satisfactory agreement with experimental data.

Notice, that similar calculations in the framework of nonlocal models of
different kind were also performed in other works (see \textit{e.g.}
\cite{Bowler:ir,DB03,Terning:1991yt,Efimov,Do0212}).

\section{Nonlocal quark-meson Lagrangian and model parameters}

The calculations are carried out in the effective chiral model with nonlocal
quark-quark interaction, which is made covariant by the inclusion of the
$P$-exponents in the non-local interaction vertex\footnote{In the description
of this section we follow \cite{RV}.}. A specific prescription for the Wilson
lines and their differentiation follows exactly
refs.~\cite{Dorokhov:2001wx,DB03}. The nonlocal $SU(2)\times SU(2)$ symmetric
quark-meson Lagrangian is given by:
\begin{align}
&  \mathcal{L}(q,\bar{q},\sigma,\pi,\rho,\omega,A)=-\frac{\pi^{a}%
(x)^{2}+\tilde{\sigma}(x)^{2}}{2G_{1}}+
  \frac{(\rho^{\mu\,a}(x))^{2}+(\omega^{\mu}(x))^{2}}{2G_{2}}+\bar
{q}(x)(i\hat{\partial}_{x}-eQ\hat{A}(x))q(x)+\label{lagr}\\
&  \,\int d^{4}x_{1}d^{4}x_{2}\,f(x-x_{1})f(x_{2}-x)\bar{q}(x_{1}%
)E(x_{1},x)\left(  \tilde{\sigma}(x)+
    \pi^{a}(x)i\gamma^{5}\tau^{a}+\rho^{\mu\,a}(x)\gamma^{\mu}%
\tau^{a}+\omega^{\mu}(x)\gamma^{\mu}\right)  E(x,x_{2})q(x_{2}),\nonumber\\
&  E(x,y)=\mathrm{Pexp}\left(  -ieQ\int\limits_{x}^{y}A^{\mu}(z)dz_{\mu
}\right)  ,\nonumber
\end{align}
where $\tau^{a}$ are the Pauli matrices and $\gamma^{\mu},\gamma^{5}$ are the
Dirac matrices; $\bar{q}(x)=(\bar{u}(x),\bar{d}(x))$ are the $u$ and $d$ quark
fields; $\tilde{\sigma}(x)$, $\pi(x)$, $\rho(x)$, $\omega(x)$ are the $\sigma
$, $\pi$, $\rho$, $\omega$ meson fields, respectively; $A^{\mu}(x)$ is the
photon field; $Q=1/2(\tau^{3}+1/3)$ is the electric charge operator, $G_{1}$,
$G_{2}$ are the dimensional constants of the four-quark interaction, and
$E(x,y)$ is the Schwinger phase factor.

The field $\tilde{\sigma}(x)$ has a nonzero vacuum expectation value
$\langle\tilde{\sigma}\rangle_{0}=\sigma_{0}\neq0$. In order to obtain a
physical scalar field with zero vacuum expectation value, it is necessary to
shift the scalar field as $\tilde{\sigma}(x)=\sigma(x)+\sigma_{0}$. In the
momentum representation this leads to the appearance of the dynamical quark
mass function $m(p)=-\sigma_{0}f^{2}(p)$. From the condition of absence of
linear $\sigma$ term in the Lagrangian (\ref{lagr}) one obtains the gap
equation for the dynamical quark mass
\begin{equation}
m(p)=f^{2}(p)G_{1}\frac{8N_{c}}{(2\pi)^{4}}\int d_{E}^{4}kf^{2}(k)\frac
{m(k)}{k^{2}+m^{2}(k)},\label{gap}%
\end{equation}
where $N_{c}$ is the number of colors. This and further equations are given in
the Euclidean space. For numeric estimates of the nonlocal effects we shall
use the momentum dependent dynamical mass defined by condition \cite{RV}
\begin{equation}
\frac{m^{2}(p)}{m^{2}(p)+p^{2}}=\exp\left(  -{p^{2}}/{\Lambda^{2}}\right)
.\label{cond}%
\end{equation}
This choice, following similar considerations given in
\cite{Bowler:ir,Dorokhov:2001wx,Efimov}, provides quark confinement. Indeed,
the expression for $m(p)$ is found to be:
\begin{equation}
m(p)=\left(  \frac{p^{2}}{\exp\left(  {p^{2}}/{\Lambda^{2}}\right)
-1}\right)  ^{1/2},\label{M(p)}%
\end{equation}
depends only on one free parameter $\Lambda$, has no any singularities in the
whole real axis and exponentially drops as $p^{2}\rightarrow\infty$ in the
Euclidean domain. From eq.(\ref{gap}) it follows that nonlocal form factors
have similar behavior providing the absence of UV divergences in the model. At
$p^{2}=0$ the mass function is equal to the cut-off parameter $\Lambda$,
$m(0)=\Lambda$, which is a specific feature of the model \cite{RV}. From the
gap equation we find the relation between four-quark coupling, $G_{1}$, and
the nonlocality parameter, $\Lambda$ ,
\begin{equation}
G_{1}=\frac{2\pi^{2}}{N_{c}}\frac{1}{\Lambda^{2}}.\label{G(L)}%
\end{equation}

The expressions for meson renormalization functions are found from
consideration of meson polarization operators. For the following calculations
we need only the values of these functions at $p^{2}=0$ where they take the
form
\begin{align}
&  g_{\pi}^{-2}(0)=\frac{N_{c}}{4\pi^{2}}\left(  \frac{3}{8}+\frac{\zeta
(3)}{2}\right)  \,,\,\qquad g_{\pi}(0)\approx3.7,\label{gpi}\\
&  g_{\rho}^{2}(0)=\frac{M_{\rho}^{2}}{G_{2}^{-1}+\Pi_{\rho}^{T}(0)}%
,\qquad\,g_{\rho}(0)=g_{\omega}(0),\label{grho}%
\end{align}
where $\zeta$ is the Riemann zeta function, $\Pi_{\rho}^{T}(p)$ is the
transversal part of the $\rho$-meson polarization operator. In the chiral
limit one has two arbitrary parameters $\Lambda$, $G_{2}$. We fix their values
with help of the weak pion decay constant $f_{\pi}=93$ MeV, and $\rho$-meson
mass $M_{\rho}=770$ MeV. By using the Goldberger-Treiman relation $g_{\pi
}(0)=m(0)/f_{\pi}$ one finds $\Lambda=m(0)=340$ MeV. The constant $G_{2}$ is
found as $G_{2}^{-1}=-\Pi_{\rho}^{T}(M_{\rho})$. Fitting this value to the
$\rho$-meson mass we obtain $G_{2}=6.5$ GeV$^{-2}$ that corresponds to the
coupling $g_{\rho}^{2}(0)\approx M_{\rho}^{2}G_{2}\approx2$.

\begin{figure}[ptb]
\hspace{0.3cm}
\resizebox{0.550\textwidth}{!}{\includegraphics{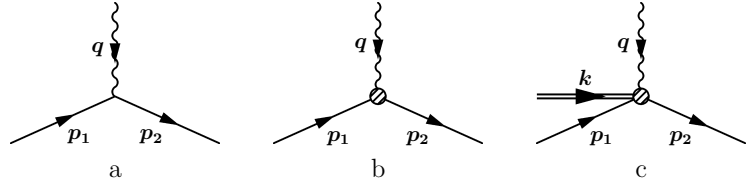}}\caption{Photon
vertices: a) the local vertex; b,c) quark-photon and quark-photon-meson
nonlocal vertices.}%
\label{photonv}%
\end{figure}

Apart from the usual local quark-photon vertex in the Lagrangian (\ref{lagr})
there appear quark-photon and quark-photon-meson nonlocal vertices generated
by $\mathrm{Pexp}$ (see fig. \ref{photonv}). 
The details of calculation of
these vertices can be found in \cite{Bowler:ir,DoLT98,DB03,Terning:1991yt}.
The explicit form of the vertices appearing in the diagrams describing
photon--vector-meson transitions (see fig. \ref{phrho}) is given in Appendix.

\section{Electromagnetic radius of the charged pion.}

\begin{figure}[ptb]
\hspace{0.3cm}
\resizebox{0.550\textwidth}{!}{\includegraphics{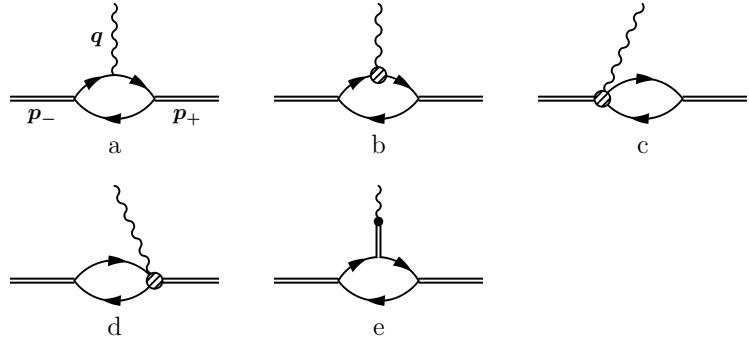}}\caption{Diagrams
describing charged pion radius: a) the local contribution; b-d) nonlocal
contributions; e) the diagrams with the $\rho$-meson.}%
\label{pchrad}%
\end{figure}

The diagrams describing the charged pion radius are presented in fig.
\ref{pchrad}. 
The quark loop diagrams with local and nonlocal photon vertices
(contact diagrams) are given in the figs. \ref{pchrad}a-d \ and the diagram
with the intermediate $\rho$-meson is drawn in fig. \ref{pchrad}e. The
amplitude for the process has the form
\begin{equation}
T_{\gamma^{\ast}\pi^{+}\pi^{-}}=e(p_{+}+p_{-})^{\mu}A^{\mu}(q)\pi^{+}%
(p_{+})\pi^{-}(p_{-})F_{\gamma^{\ast}\pi^{+}\pi^{-}}(q^{2}),
\end{equation}
where $F_{\gamma^{\ast}\pi^{+}\pi^{-}}(q^{2})$ is pion form-factor and
$q=p_{+}-p_{-}$. The electromagnetic pion radius $\langle r^{2}\rangle
_{\mathrm{em}}$ is defined by
\[
\langle r^{2}\rangle_{\mathrm{em}}=-6\left.  \frac{dF_{\gamma^{\ast}\pi^{+}%
\pi^{-}}(q^{2})}{dq^{2}}\right|  _{q^{2}=0}.
\]

In the local NJL model the contact diagram gives about $80\%$ of the correct
value for the pion radius \cite{Gerasimov,Volkov:1996rc}. Therefore, taking
into account the diagram with intermediate $\rho$-meson \cite{Volkov:1996rc}
leads to a too large value for the pion radius in comparison with the
experimental value \cite{Hagiwara:fs}. Indeed, in the local NJL model one has
\begin{align}
\langle r^{2}\rangle_{\mathrm{cont}}^{\mathrm{NJL}} &  =\frac{N_{c}}{4\pi^{2}f_{\pi}^{2}%
}=0.342\,\mathrm{fm}^{2},\nonumber\\
\langle r^{2}\rangle_{\rho}^{\mathrm{NJL}} &  ={6}/{M_{\rho}^{2}}=0.394\,\mathrm{fm}%
^{2},\nonumber\\
\langle r^{2}\rangle_{\mathrm{em}}^{\mathrm{NJL}} &  = 0.736\,\mathrm{fm}^{2},%
\label{RnjlEM}\\
\langle r^{2}\rangle_{\mathrm{exp}} &  =0.451\pm0.011\,\mathrm{fm}^{2}.%
\label{RexpEM}%
\end{align}

Now we show that in contrast, in the nonlocal models the diagram with the
intermediate vector-meson is noticeably suppressed. In the model considered
the contributions to the e.m. pion radius from diagrams in figs. \ref{pchrad}%
a-d and from diagram in fig. \ref{pchrad}e equal $\langle r^{2}\rangle
_{\mathrm{cont}}=0.340$ fm$^{2}$ and $\langle r^{2}\rangle_{\rho}=0.047$
fm$^{2}$, respectively. Then the e.m. pion radius becomes $\langle
r^{2}\rangle_{\mathrm{em}}=0.387$ fm$^{2}$, that is in much better agreement with
experimental value than the result of the local model.

Let us consider in more details the contribution of the $\rho$-meson diagrams
in both the local NJL model and the nonlocal model. These diagrams consist of
three parts: the photon--$\rho$-meson transition, the $\rho$-meson propagator,
and the part describing the $\rho\rightarrow\pi\pi$ vertex. Our calculations
show that in the nonlocal quark models the first part is more than twice
smaller than in NJL, the second part does not
changed, and the third part is four times smaller than in the NJL model. The
latter suppression is due to strong $q^{2}$ dependence of the amplitude
occurring in the nonlocal models. As a result, we obtain almost one order
decrease of the $\rho$-meson diagram contribution in the nonlocal models with
respect to the prediction of the local NJL model.

\begin{figure}[ptb]
\hspace{0.3cm} \resizebox{0.53\textwidth}{!}{
\includegraphics{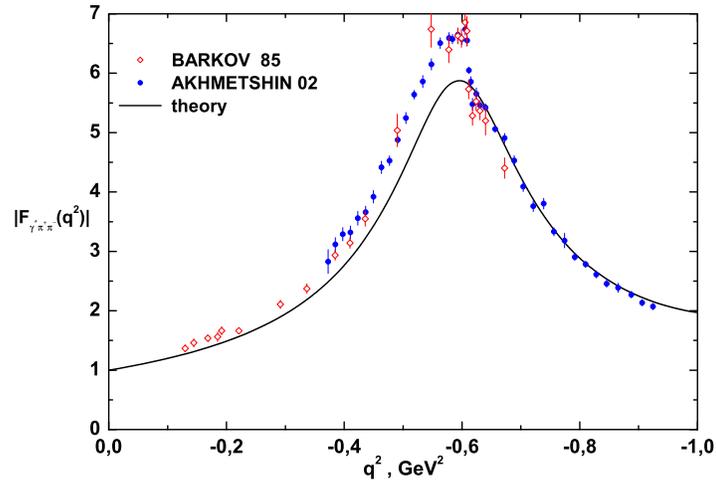}
}\caption{The absolute value of charged pion form-factor in the 
time-like region. The
finite width of $\rho$-meson is $\Gamma_{\rho}=135$ MeV \cite{RV}.
Experimental data are taken from \cite{Barkov:ac},\cite{Akhmetshin:2001ig}. }%
\label{fppgt}%
\end{figure}

\begin{figure}[ptb]
\hspace{0.3cm} \resizebox{0.53\textwidth}{!}{
\includegraphics{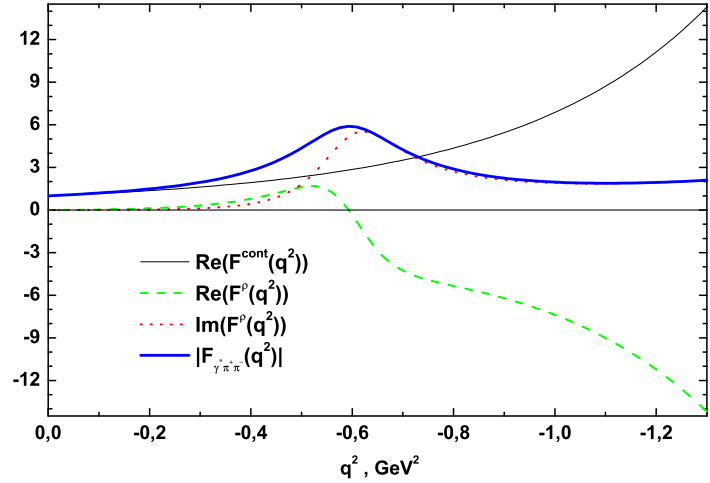}
}\caption{Partial contributions to the charged pion form-factor in the 
time-like region from the
contact and $\rho$-meson diagrams and the absolute value of
the charged pion form-factor($\mathrm{Im(F^{cont}(q^2))=0}$).}%
\label{fppgtpartial}%
\end{figure}

\begin{figure}[ptb]
\hspace{0.3cm} \resizebox{0.53\textwidth}{!}{
\includegraphics{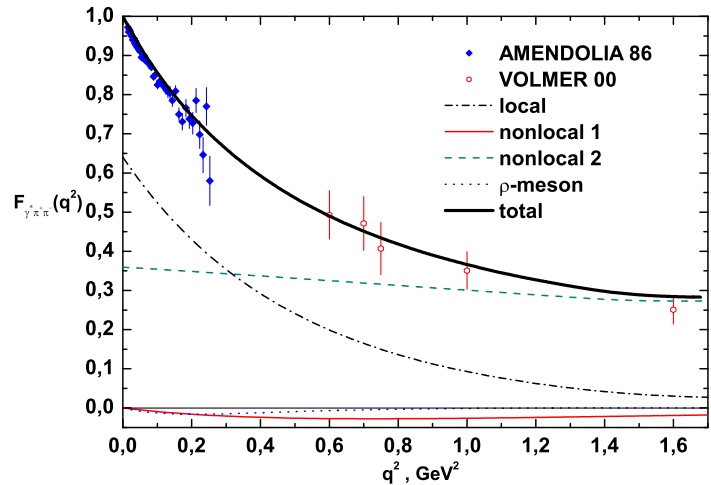}
}\caption{The charged pion form-factor in the space-like region. 
Partial contributions
from diagrams fig.\ref{pchrad}a(local), fig. \ref{pchrad}b(nonlocal 1),
fig.\ref{pchrad}c,d(nonlocal 2), and fig.\ref{pchrad}e($\rho$-meson) are shown.
Experimental data are taken from \cite{Amendolia:1986wj}, \cite{Volmer:2000ek}%
. }%
\label{fppgs}%
\end{figure}

The charged pion form-factors $F_{\gamma^{\ast}\pi^{+}\pi^{-}}(q^{2})$ in
time-like and space-like regions are shown in figs. \ref{fppgt},
\ref{fppgtpartial}, \ref{fppgs}. 
The $\rho$-meson resonance is displayed in the
time-like domain near $-q^{2}\approx M_{\rho}^{2}$ and we take into account
the $\rho$-meson decay width.

\section{Transition radius of neutral pion.}

\begin{figure}[ptb]
\hspace{0.3cm}
\resizebox{0.550\textwidth}{!}{\includegraphics{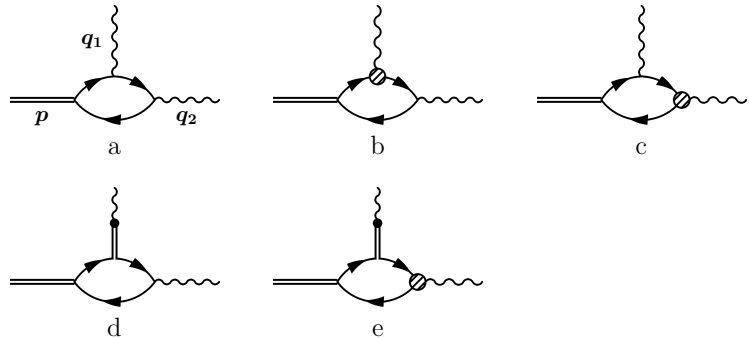}}\caption{Diagrams
describing the transition neutral pion radius: a) the local contribution; 
b-c) nonlocal contributions; d,e) the diagrams with vector mesons.}%
\label{ptrrad}%
\end{figure}

The process $\pi^{0}\rightarrow\gamma^{\ast}\gamma$ is described by the
diagrams in fig. \ref{ptrrad}. 
Figs. \ref{ptrrad}a-c correspond to the quark
loop diagrams with the local and nonlocal photon vertices (contact diagrams),
whereas figs. \ref{ptrrad}d,e correspond to diagrams with intermediate vector
meson. The amplitude of the process has the form
\begin{equation}
T_{\pi_{0}\gamma\gamma^{\ast}}=\frac{e^{2}}{4\pi^{2}f_{\pi}}\epsilon^{\mu
\nu\alpha\beta}q_{1}^{\nu}q_{2}^{\mu}A^{\alpha}(q_{1})A^{\beta}(q_{2})\pi
^{0}(p)F_{\pi_{0}\gamma\gamma^{\ast}}(q_{1}^{2}),\label{tpgg}%
\end{equation}
where $q$, $q_{2}$, $p$ are the two photon and the pion momentum,
respectively. Transition pion radius is defined as \footnote{The value of
transition pion radius $\langle r^{2}\rangle_{\pi_{0}\gamma^{\ast}\gamma
^{\ast}}$ does not depend on asymmetry $\omega=2(q_{1}^{2}-q_{2}^{2}%
)/(q_{1}^{2}+q_{2}^{2})$ of photon virtualities $q_{1}^{2}$ and $q_{2}^{2}$.
In the case of both off-shell photons derivative in the definition of radius
is taken with respect to the total virtuality of photons.}
\begin{equation}
\langle r^{2}\rangle_{\pi_{0}\gamma\gamma^{\ast}}=-6\left.\frac{dF_{\pi_{0}%
\gamma\gamma^{\ast}}(q_{1}^{2})}{dq_{1}^{2}}\right|_{q_{1}^{2}=0}.\label{Rtr}%
\end{equation}

In the local NJL model the contact diagrams give 
$\langle r^{2}\rangle_{\mathrm{cont}}^{\mathrm{NJL}}={1}/{2m^{2}}$ \cite{Gerasimov},
where $m$ is the constituent quark mass.
Then in the NJL model without and with the $\pi-a_{1}$ transition
one gets $\langle r^{2}\rangle_{\mathrm{cont}}^{\mathrm{NJL}}=0.342$
fm$^{2}$ for $m=240$ MeV and $\langle r^{2}\rangle_{\mathrm{cont}}^{\mathrm{NJL}}=0.248$
fm$^{2}$ for $m=280$ MeV \cite{Volkov:zb}, respectively. The experimental value of the
transition pion radius is \cite{Hagiwara:fs}
\begin{equation}
\langle r^{2}\rangle_{\pi_{0}\gamma\gamma^{\ast}}^{\exp}=0.407\pm0.051\quad
\mathrm{fm}^{2}.\label{RtrExp}%
\end{equation}
However in the local model, the diagrams with vector mesons, figs.
\ref{ptrrad}d,e, again additionally contribute as \\
$\langle r^{2}\rangle_{\rho+\omega}^{\mathrm{NJL}}={6}/{M_{\rho}^{2}}=0.394$ fm$^{2}$
that noticeably enlarges the
value of the transition radius in comparison with experiment.

In contrast, in the nonlocal models the vector meson diagrams are strongly
suppressed. Indeed, in the model considered one has $\langle r^{2}%
\rangle_{\mathrm{cont}}=0.308$ fm$^{2}$ and $\langle r^{2}\rangle_{\rho
+\omega}=0.07 $ fm$^{2}$, respectively. Then, we obtain $\langle r^{2}%
\rangle_{\pi_{0}\gamma\gamma^{\ast}}=0.378$ fm$^{2}$ that is in satisfactory
agreement with experimental data.

We have to note that in the local NJL model the diagram $\rho\rightarrow
\pi\gamma$ provides $g_{\rho}/2\approx3$ (in normalization of eq.
(\ref{tpgg})) and in the nonlocal model it gives $\approx1.2$. Taking into
account the value of photon--vector-meson couplings, which are the same as in
the case of charged pion, we see that in the nonlocal models the contribution
of vector mesons is more than four times smaller than in the local model.

\section{Conclusion}

We would like to emphasize that nonlocal models allow us to solve a set of
problems that can not be solved in the local NJL models. First of all it
concerns the correct description of momentum dependence of meson amplitudes
in the low-energy domain. One of the examples of that is the calculation of
electromagnetic meson radii. This problem is considered in present work.

Our calculation is devoted to the investigation of two subjects. First, the
region of the effective quark interaction inside the pion is determined. The
size of this region corresponds to the electromagnetic and transition radii of
the pion. Here, we have shown that in the model considered this region is in
agreement with experimental data.

Next, we study relative contributions to the pion form-factor $F_{\gamma
^{\ast}\pi^{+}\pi^{-}}$ and pion radii from the contact diagrams and diagrams
with vector mesons as intermediate states. We compare these contributions to
the pion radii with the results obtained in the framework of the local NJL
model. We show that in the nonlocal models the contribution of vector mesons
is noticeably smaller than that of the contact diagrams in contrast with the
local NJL model where these contributions are comparable. It is worth to note
that in the nonlocal models the additional diagrams with photon interacting
nonlocally with quark appear. These diagrams are important for gauge
invariance. The contributions of nonlocal contact diagrams to the pion radii
have the same order of magnitude as the contributions of the diagrams with
vector mesons\footnote{The contribution of contact diagrams in nonlocal
models is comparable with that of local models.}. 
Our calculations show that the diagram with local quark-photon vertex 
gives the dominant part of the pion radii(see table \ref{table1}).

We have to note finally that one expects further corrections to the pion radii
resulted from the so-called $1/N_{c}$ corrections that are presumably of the
same order as the vector meson corrections considered in the present work (see
also \cite{Hippe:hu}).

The vector-meson diagrams play a very important role in the description of the
pion form-factor $F_{\gamma^{\ast}\pi^{+}\pi^{-}}$ in the time-like region.
These diagrams allow us not only to describe the $\rho$-meson resonance but
also to obtain a correct behavior of the form-factor in the region
$-q^{2}>M_{\rho}^{2}$. It should be noted that the contact and vector meson
diagrams increase in absolute values in the region $-q^{2}>M_{\rho}^{2}$, but
they have opposite signs. As a result, the total contribution decreases in
(see fig. \ref{fppgtpartial})
agreement with experimental tendency \cite{Akhmetshin:2001ig}.

Similar situation is observed in the nonlocal model considered in
\cite{Bowler:ir}. However, in the present model there is an additional relation
between model parameters $m(0)$ and $\Lambda$, $m(0)=\Lambda$, governed by
eq.(\ref{cond}). As a result, arbitrary parameters are absent in contrast to
\cite{Bowler:ir}. Nevertheless, in our model the fine-tuning that leads to
cancellation of different contributions occurs automatically in accordance
with experimental data.

In future, we plan to calculate pion electric and magnetic
polarizabilities and $\pi$--$\pi$ scattering lengths.

We are grateful to V.~V.~Anisovich for drawing our attention to the subject of
this work and S.~B.~Gerasimov for useful discussions. The work is supported by
RFBR Grant no. 02-02-16194 and the Heisenberg--Landau program.
AED thanks for partial support from RFBR (Grants nos.
01-02-16431, 03-02-17291) and INTAS (Grant no. 00-00-366).

\section{Appendix}

Here we present the vertices used in the above calculations. The nonlocal
quark-photon vertex is:
\begin{equation}
-eQ(p_{1}+p_{2})^{\mu}\frac{m(p_{1})-m(p_{2})}{p_{1}^{2}-p_{2}^{2}}\bar
{q}(p_{1})q(p_{2})A_{\mu}(p_{1}-p_{2}),
\end{equation}
while the nonlocal quark-photon-meson vertex takes the form
\begin{align}
&  e \bar{q}(p_{1})q(p_{2})A_{\mu}(q) g_{M}(k) M(k)
  \left(  \frac{f(p_{1})-f(p_{1}+q)}{p_{1}^{2}-(p_{1}+q)^{2}}f(p_{2}%
)(2p_{1}+q)^{\mu}Q\Gamma_{M}+\right. 
  \left.  \frac{f(p_{2})-f(p_{2}-q)}{p_{2}^{2}-(p_{2}-q)^{2}}f(p_{1}%
)(2p_{2}-q)^{\mu}\Gamma_{M}Q\right)  ,
\end{align}
where $q$, $k$, $p_{1}$, $p_{2}=p_{1}+q+k$ are the momenta of the photon,
meson, antiquark and quark, respectively; $M$ is meson field; $g_{M}$ is the
function describing renormalization of meson field; the matrices $\Gamma_{M} $
are defined by
\[
\Gamma_{\pi}^{a}=i\gamma^{5}\tau^{a},\quad\Gamma_{\rho}^{\mu\,a}=\gamma^{\mu
}\tau^{a},\quad\Gamma_{\omega}^{\mu}=\gamma^{\mu}.
\]

\clearpage

\begin{table}

\begin{tabular}{|c|c|c|c|c|}\hline
contribution, fm$^{2}$
&
$\langle r^{2}\rangle_{\mathrm{em}}$ &
$\langle r^{2}\rangle_{\pi_{0}\gamma\gamma^{\ast}}$&
$\langle r^{2}\rangle_{\mathrm{em}}^{\mathrm{NJL}}$&
$\langle r^{2}\rangle_{\pi_{0}\gamma\gamma^{\ast}}^{\mathrm{NJL}}$\\\hline
local & 0.302 & 0.243 & 0.342 & 0.342(0.248)\\\hline
nonlocal & 0.038 & 0.065 &  & \\\hline
vector & 0.047 & 0.07 & 0.394 & 0.394\\\hline
total & 0.387 & 0.378 & 0.736 & 0.736(0.642)\\\hline
\end{tabular}
\caption{The comparison of the pion radii obtained in the nonlocal
and local quark models. The local contribution is that from diagrams with
local photon vertex (fig. \ref{pchrad}a and fig. \ref{ptrrad}a). 
The nonlocal contribution is that
from diagrams with additional nonlocal photon vertex (figs. \ref{pchrad}b-d 
and figs. \ref{ptrrad}b,c). The vector-meson contributions are that from 
diagrams with intermediate vector mesons(fig. \ref{pchrad}e 
and figs. \ref{ptrrad}d,e).}
\label{table1}
\end{table}

\begin{figure}[ptb]
\hspace{0.3cm}
\resizebox{0.550\textwidth}{!}{\includegraphics{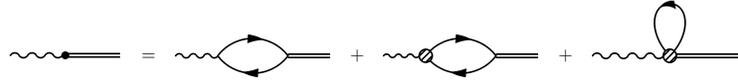}}\caption{Diagrams
describing vector-meson--photon mixing.}%
\label{phrho}%
\end{figure}

\end{document}